\documentclass[a4paper]{jpconf}
\usepackage{graphicx}
\usepackage{dcolumn}
\usepackage{bm}
\usepackage{units}
\usepackage{upgreek}
\usepackage{braket}
\usepackage{ucs}
\usepackage{color}
\usepackage{layouts}
\usepackage{hyperref}
\usepackage[all]{hypcap}
\usepackage{standalone}
\usepackage{tikz}
\usepackage{pgf}
\usepackage[utf8x]{inputenc}
\usepackage{subfigure}

\begin{document}

\title{Dynamically Probing Ultracold Lattice Gases via Rydberg Molecules} 

\author{T. Manthey, T. Niederpr\"um, O. Thomas, H. Ott}
\ead{ott@physik.uni-kl.de}
\address{Department of Physics and Research Center OPTIMAS, Technische Universität Kaiserslautern, Gottlieb-Daimler-str. 46 67663 Kaiserslautern, Germany}

\begin{abstract}
We show that the excitation of long-range Rydberg molecules in a three-dimensional optical lattice can be used as a position- and time-sensitive probe of the site occupancy in the system. To this end, we detect the ions which are continuously generated by the decay of the formed Rydberg molecules. While a superfluid gas shows molecule formation for all parameters, a Mott insulator with $n=1$ filling reveals a strong suppression of the number of formed molecules. In the limit of weak probing, the technique can be used to probe the superfluid to Mott-insulator transition in real-time. Our method can be extended to higher fillings and has various applications for the real-time diagnosis and manipulation of ultracold lattice gases. 
\end{abstract}


The quantum phase transition from a superfluid to a Mott insulator in ultracold atomic lattice gases \cite{Mott1974,Greiner2002,Joerdens2008} highlights the large potential of atomic quantum gases for the study of many-body quantum systems. One of the most prominent signatures of the Mott insulating phase is the suppression of atom number fluctuations. Measuring the site occupancy has been a key instrument to diagnose the properties of these systems. The employed techniques include the measurement of doubly occupied lattice sites with spin changing collisions \cite{Gerbier2006,Foelling2006}, imaging the Mott shells via microwave clock states \cite{Campbell2006} or direct \textit{in situ} fluorescence imaging \cite{Sherson2010,Bakr2010}. In all these approaches, the measurement is connected with the destruction of the whole atomic sample. Here, we introduce a new technique based on the formation of long-range Rydberg molecules in an optical lattice. In the limit of weak probing, the technique allows for a real-time detection of the site occupancy.

Rydberg atoms feature prominent properties such as a giant dipole moment, which allows them to strongly interact with other excited atoms over distances of several micrometers \cite{Beguin2013}. This leads to the so-called Rydberg blockade \cite{Tong2004}, which prevents the excitation of more than one atom in a micrometer-sized volume \cite{Gaetan2009,Urban2009,Weber2015}. Another peculiar feature is that the electron in a Rydberg state has a contact interaction with ground state atoms immersed in the electron's wave function. For an attractive interaction, this leads to the formation of long-range Rydberg molecules \cite{Greene2000,Bendkowsky2009,Li2011}. In order to make this process happen, two ground state atoms have to be closer to each other than the extension of the electronic wave function of the Rydberg electron, which is in the order of 100 - 1000\,nm for main principal quantum numbers between $n=30$ to $n=100$. In an optical lattice configuration in the tight binding limit, the parameters can be chosen such that only two atoms at the same lattice site can form a molecule. The formation of molecules is then proportional to the probability to find two atoms at the same site. This can be used as a diagnostic tool for the site occupancy. 

\begin{figure}
	\centering
		\includegraphics[scale=1.0]{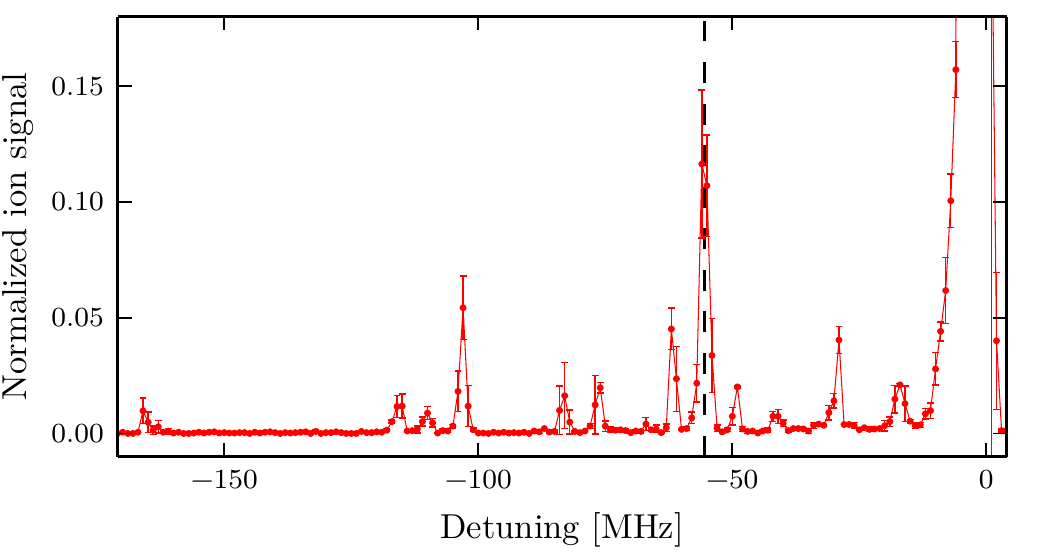} 
	\caption{Spectrum of Rydberg molecules for the excitation to the $30P_{3/2}$ state in $^{87}$Rb. The measurement was done with a Bose-Einstein condensate in the $\ket{F=1,m_F=1}$ state at a residual magnetic field of approximately $\unit[1]{G}$. The signal is normalized to the ion yield on resonance. Throughout this work, the measurements were taken on the molecular peak around $-55$\,MHz, indicated with the dashed line. The red solid line is a guide to the eye.}
	\label{fig:fig1_spectroscopy}
\end{figure}

In the experiment, we use an ultracold gas of rubidium 87 atoms in a crossed optical dipole trap with equal trapping frequencies ($\omega=2\pi\times 65 s^{-1}$) in all three directions \cite{Manthey2013}. The atoms can additionally be loaded in a superimposed three-dimensional optical lattice with a wavelength of $\lambda =\unit[742]{nm}$. Two axes are realized in a retro reflected configuration, while in the third axis -- due to geometrical contraints -- two orthogonal laser beams are used. This leads to lattice constants of $d_1 = d_2= \unit[371]{nm}$ and $d_3 = \unit[524]{nm}$. We excite the atoms to the $30P_{3/2}$ state using a UV laser at \unit[297]{nm}. The UV light is produced by a frequency doubled dye laser locked to a transfer cavity \cite{Manthey2013,Thoumany2009}. The choice of the $30P_{3/2}$ state is a compromise between a sufficiently small electronic wave function and reliable laser operation. After excitation, the molecules can decay into molecular ions \cite{Niederpruem2015}, can be photoionized by the trapping lasers and black body radiation \cite{Beterov2009}, resulting in the production of atomic ions, or can undergo a spontaneous emission. The two ion signals are proportional to the probability of having created a molecule at a lattice site, thus allowing for dynamically probing the site occupancy.

In Fig.\,\ref{fig:fig1_spectroscopy} we show the spectrum of Rydberg molecules for the $30P_{3/2}$ state of rubidium. The measurement was performed in a Bose-Einstein condensate (BEC) in the $\ket{F=1,m_F=1}$, which was continuously excited for \unit[20]{ms} (P=$\unit[3]{\mu W}$). The spectrum shows several molecular peaks red-detuned from the main resonance. The peak at a detuning of about $-55$\,MHz from resonance is chosen for the experiments presented in this work. A scaling analysis of this peak with varying atom number (see also later on) reveals that this peak corresponds to a diatomic Rydberg molecule. The large detuning from resonance ensures that the off-resonant excitation of single atoms is highly suppressed. The extension of a Rydberg molecule is given by the diameter of the electronic wave function. For $n=30$ this amounts to $d_\mathrm{Ry} = \unit[146]{nm}$, which is significantly smaller than the lattice spacing. As the molecule formation requires a contact interaction between the Rydberg electron and a ground state atom, molecule formation can only take place on the same lattice site. Compared to the Wannier function in the tight binding limit (the $\sigma$-widths of the Gaussian onsite density distribution for $s=20$ are $\sigma_1=\sigma_2=60\,nm$ and $\sigma_3=85\,nm$ for the different lattice axis) are of similar magnitude.

\begin{figure}
	\centering
		\includegraphics[scale=1]{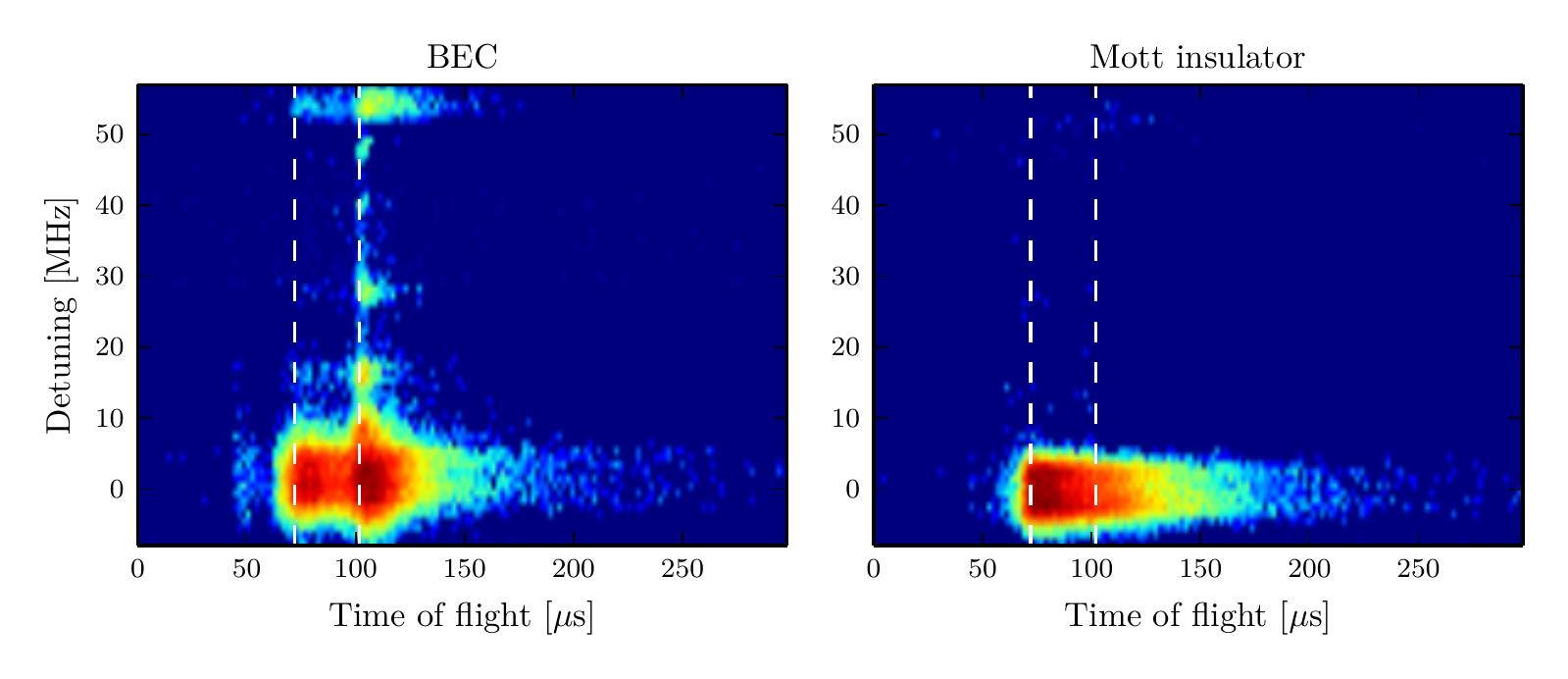} 
	\caption{Time of flight spectra for a BEC and a Mott insulator for varying detuning. The atom number was reduced, such that the $n=2$ shell in the Mott insulator is suppressed. The data was averaged over 2000 pulses with $\unit[1]{\mu s}$ length. The signal starting at $t=70\,\mu$s stems from atomic ions (left dashed line), while the signal starting at $t=100\mu s$ originates from molecular ions (right dashed line).}
	\label{fig:fig2_TOF_detuning}
\end{figure}

The sensitivity to doubly occupied sites can be directly proven by comparing the molecular spectrum of a BEC with that of a Mott insulator with a filling of $n=1$ in the trap center. The latter is realized by adiabatically ramping up the three dimensional optical lattice with a lattice depth of $20 E_\mathrm{rec}$ in all three directions. In Fig. \ref{fig:fig2_TOF_detuning} we compare the two spectra for a small system, where the $n=2$ Mott shell is suppressed. To distinguish the produced atomic ions from molecular ions we have measured the spectrum in a time of flight setting, where the sample is excited with short laser pulses of $\unit[1]{\mu s}$ duration ($\unit[10]{\mu W}$ power). For a BEC, atomic ions as well as molecular ions are created both, on resonance and for red detuning. The molecular peaks for finite detuning correspond to the ones shown in Fig.\,\ref{fig:fig1_spectroscopy}. The generation of atomic ions on the molecular resonances is caused by direct photoionization of the Rydberg molecule, while the molecular ions are created by associative ionization \cite{Niederpruem2015}. In case of the Mott insulator, no excitations of molecules are seen, and only atomic ions are created on resonance. This demonstrates that Rydberg molecule formation is a sensitive probe for studying the site occupancy in optical lattices.

\begin{figure}
	\centering
		\includegraphics[scale=1]{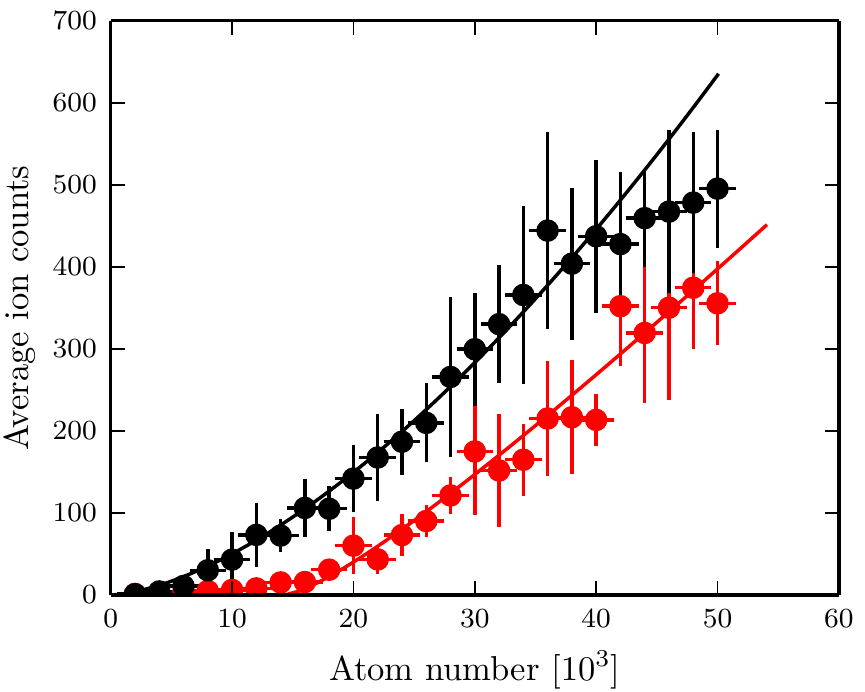} 
	\caption{Number of produced ions in dependence of the total atom number $N$ for a BEC (black points) and Mott insulator (red points). The sample was continuously excited for $\unit[20]{ms}$. The lines are fits as explained in the text.}
	\label{fig:fig3_molpeak1_atomnumber_dependency}
\end{figure}

The method can now be used to measure the critical atom number $N_\mathrm{crit}$ for the transition to the $n=2$ Mott shell. To this end, we vary the atom number of the sample before we ramp into the Mott insulator and excite the Rydberg molecules. In Fig.\,\ref{fig:fig3_molpeak1_atomnumber_dependency} we show the total number of detected ions during an excitation of $\unit[20]{ms}$ duration for a BEC and a Mott insulator (lattice depth 20 $E_\mathrm{rec}$) in dependency of the atom number $N$. Since the excitation of a Rydberg molecule is a two-body process, the ion production rate in a BEC, $\Gamma_\mathrm{BEC}$, scales with the volume integrated squared density. The central density in a BEC with parabolic confinement scales as $n_0 \propto N^{2/5}$ and we find after some algebra $\Gamma_\mathrm{BEC} \propto N^{7/5}$. From a power law fit to the data we extract an exponent $b=1.58 \pm 0.24$, compatible with the predicted value of $1.4$. In the case of a Mott insulator the ion production rate is proportional to the number of doubly occupied lattice sites. For an atom number which does not allow for a $n=2$ shell to build up, no molecules should be formed (compare Fig.\,\ref{fig:fig2_TOF_detuning}b). Once the atom number exceeds a critical atom number $N_\mathrm{crit}$, the $n=2$ shell starts to grow. From an analysis of the shell sizes of a Mott insulator in the atomic limit for zero temperature, one can show that the number of doubly occupied sites $Z_2$ is connected to the total number of atoms as $N=(N_\mathrm{crit}^{2/3}+Z_2^{2/3})^{3/2}+Z_2$. The number of molecules is then proportional to $Z_2$. A fit to the data (see Fig.\,\ref{fig:fig3_molpeak1_atomnumber_dependency}) with this function shows good agreement and reveals a critical atom number of $\unit[14\pm 3]{k}$, above which the $n=2$ shell grows. An estimate of the critical atom number in the atomic limit for $T=0$ yields $N_\mathrm{crit} =\unit[(27\pm 5)]{k}$. We attribute the difference to the finite temperature and the finite tunneling coupling, which both lead to the formation of doubly occupied sites for smaller total atom numbers \cite{Gerbier2007}.

\begin{figure}
	\centering
		\includegraphics[scale=1]{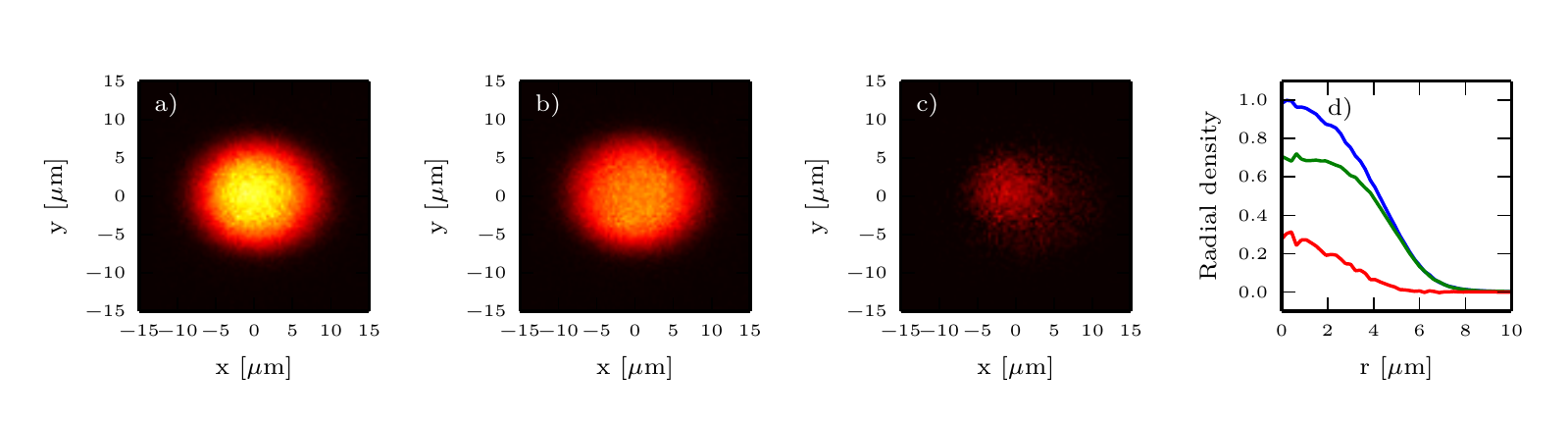} 
	\caption{Electron microscope image of a Mott insulator before (a)  and after
 (b) an illumination with 200 short excitation pulses ($P = \unit[1]{mW}, \tau= \unit[1]{\mu s}$). (c) Difference between both signals, highlighting the spatial regions where atoms have been lost from the system. (d) Radial density for (a)-(c) (blue: without UV pulse, green: with UV pulse, red: difference).}
	\label{fig:fig4_spacial_density_UV_irradiation}
\end{figure}

When a molecule is excited in the lattice, several inelastic processes occur. Besides the decay into a molecular ion, photoionization and spontaneous decay are also possible. In most cases, it will be rather unlikely that after the decay both atoms are again in the ground state of the lattice site. The excitation of a molecule is therefore connected with a loss of at least one of the atoms. This gives a handle to the spatially resolved measurement of the site occupancy in the lattice. With a scanning electron microscope \cite{Gericke2008,Santra2015} that is implemented in our setup, it is possible to image this change of density. This is done by making an image of the Mott insulator with and without a UV excitation. Fig. \ref{fig:fig4_spacial_density_UV_irradiation}a) and b) show both density profiles and the difference of both signals is shown in Fig. \ref{fig:fig4_spacial_density_UV_irradiation}c). The density in the center is clearly reduced compared to the wings. As losses are only expected to appear in the $n=2$ shell, the radius of the loss area should indicate the transition to the $n=1$ shell. In the experiment we find a radius of $6\pm1\,\mu$m. The estimated radius of $7.5\,\mu$m is indeed close to this value.

An important feature of the excitation to molecular Rydberg states in a optical lattice compared to other techniques, is that the ion signal is continuously generated. If the power of the excitation laser is reduced, the sample is only slightly disturbed and a continuous measurement of the site occupancy becomes possible, even when the lattice parameters are changed. To demonstrate this, we ramp the optical lattice to $20 E_\mathrm{rec}$ within 150\,ms, while continuously exciting Rydberg molecules at a very small laser power of $P=\unit[1.5]{\mu W}$. 

\begin{figure}
	\centering
		\includegraphics{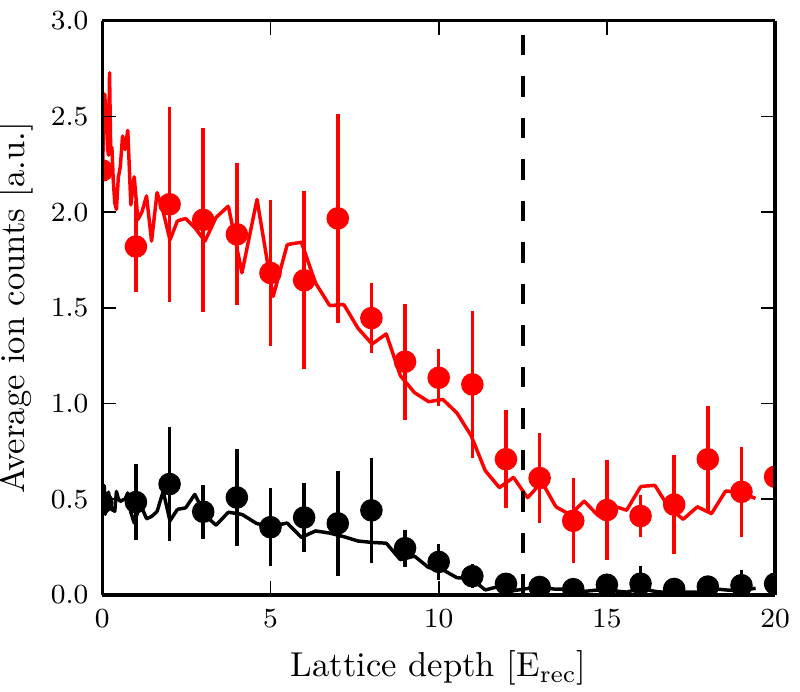}
	\caption{Molecular signal in dependence of the lattice depth for an atom number above (red) and below (black) $N_\mathrm{crit}$. The dots represent the data taken in a static configuration, where the system was probed for 20\,ms after an adiabatic ramp in the lattice. The lines represent the data taken in a dynamic configuration, where the system is probed during a 150\,ms ramp. The red line has been multiplied by a factor of 1.5 (see text). The dashed vertical line indicates where the superfluid-Mott-transition takes place.}
	\label{fig:fig5_Sdep_dynamic_static}
\end{figure}

The time-resolved signal is shown in Fig.\,\ref{fig:fig5_Sdep_dynamic_static}, where the time axis has been converted to the actual lattice depth. We compare the dynamic measurement with a static measurement where the system was adiabatically prepared for a given lattice depth and probed for $\tau= \unit[20]{ms}$. The measurement was performed for an atom number below and above $N_\mathrm{crit}$. In the case $N_{tot}>N_{crit}$ the ion rates from the dynamic case a factor of $1.5$ smaller than in the static one. This is most likely due to shifts of the molecular resonance from shot to shot because of electric stray fields in the experimental chamber. Apart from this normalization factor, the shape of the signal is identical. This demonstrates that the excitation to Rydberg molecules allows for a time-resolved analysis of the site occupancy in an optical lattices. In future experiments, this can be used to study the dynamics of doublon formation with high temporal resolution. 

We have demonstrated that the excitation of Rydberg molecules in an optical lattice is a valuable tool to study the on-site occupancy in static and dynamic configurations. Due to the imprinted losses, the approach is position sensitive, provided it is combined with a spatially resolved \textit{in situ} imaging method. Various intriguing applications of this approach are feasible. First, the technique is applicable to any bosonic or fermionic species, where long-range Rydberg molecules can be excited. Also mixtures of different hyperfine states or even different atomic species can be probed. Second, for high principal quantum numbers, where the electronic wave function reaches to the neighboring lattice site, nearest neighbor density correlations could be probed. Eventually, the existence of Rydberg molecules including more than one ground state atom allows the probing of multiply occupied lattice sites. 

We acknowledge financial support by the DFG within the SFB/TR49. O.T is funded by the graduate school of excellence MAINZ.\\

\section*{References}

\bibliography{Motttransition_ref}

\end{document}